\documentclass[sigconf,authorversion]{acmart}
\AtBeginDocument{%
  }

\usepackage{subcaption}

\setcopyright{cc}
\setcctype{by}
\copyrightyear{2026}
\acmYear{2026}
\acmConference[MuC'26]{Mensch und Computer 2026 -- Tagungsband, Gesellschaft f\"ur Informatik e.V.}{30. August - 02. September 2026}{Duisburg, Germany}
\acmDOI{10.18420/muc2026-mci-wip-344}
\begin{document}

\title{Transforming Interactions in Thesis Supervision: An Expos\'e-First Workflow in Higher Education}

\author{Lin-Yin Huang}
\orcid{0009-0005-7637-3262}
\affiliation{%
  \institution{Ubiquitous Knowledge Processing Lab (UKP)}
  \institution{Technical University of Darmstadt}
  \city{Darmstadt}
  \country{Germany}
}
\email{lin-yin.huang@stud.tu-darmstadt.de}

\author{Dennis Zyska}
\orcid{0009-0003-4914-1015}
\affiliation{%
  \institution{Ubiquitous Knowledge Processing Lab (UKP)}
  \institution{Technical University of Darmstadt}
  \city{Darmstadt}
  \country{Germany}
}
\email{dennis.zyska@tu-darmstadt.de}

\author{Iryna Gurevych}
\orcid{0000-0003-2187-7621}
\affiliation{%
  \institution{Ubiquitous Knowledge Processing Lab (UKP)}
  \institution{Technical University of Darmstadt}
  \city{Darmstadt}
  \country{Germany}
}
\email{iryna.gurevych@tu-darmstadt.de}

\renewcommand{\shortauthors}{Huang et al.}

\begin{abstract}
At the studied research institute, one professorship oversees approximately 20 theses per semester, while day-to-day supervision is distributed among doctoral and postdoctoral researchers. To manage this supervision demand, the institute uses an exposé-first workflow in which students prepare a research proposal before entering the main thesis-writing phase. This paper asks how students, supervisors, and administrators experience the exposé-first workflow as a structured process for early thesis preparation, and how it redistributes responsibility, supervision, and administrative coordination work across roles and two digital platforms. Based on a mixed-methods study\footnote{The institutional IRB has approved the related proposal.} analyzed through Frauenberger et al.'s four reflective design lenses, the findings show that the exposé-first model made thesis preparation more structured by turning early research planning into a staged process of proposal writing, feedback, and approval. Students reported that this process helped them clarify research goals and take ownership of their research plans at an early stage. However, the workflow redistributed rather than reduced work: supervisors shifted toward iterative feedback, feasibility checking, and preliminary quality assurance, while administrators carried much of the coordination across platforms, deadlines, submissions, and feedback. The paper contributes an analysis of exposé-first thesis preparation as a sociotechnical workflow, showing how workflow redesign can improve structure while leaving essential administrative coordination work underrecognized.
\end{abstract}



\maketitle

\begin{figure*}[t]
  \centering
  \includegraphics[width=\textwidth]{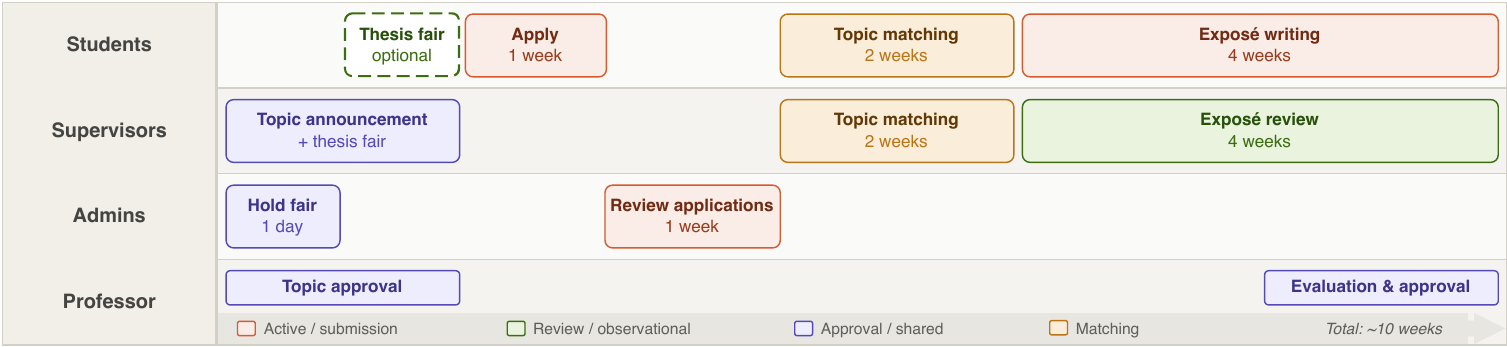}
  \caption{Overview of the expos\'e-first thesis workflow. The figure shows the main workflow phases, their approximate duration, and their distribution across students, supervisors, administrators, and the professor.}
  \Description{A workflow diagram showing topic announcement, thesis fair, application, topic matching, expos\'e writing, and final approval across students, supervisors, administrators, and the professor.}
  \label{fig:workflow-overview}
\end{figure*}

\section{Introduction}
Higher education has seen sustained enrollment growth in recent years~\cite{unesco2025higher}, increasing the demand for thesis supervision within research institutes. At the studied institute, one professorship oversees approximately 20 theses per semester, while doctoral and postdoctoral researchers provide day-to-day supervision. In this setting, thesis preparation depends on coordinated handoffs between students, supervisors, administrators, and the professor. Without a standardized process, topic matching, early research planning, feedback, and approval are difficult to align across many parallel thesis projects.

Prior research shows that clear guidance, timely feedback, and agreed expectations are central to students' progress and satisfaction during thesis supervision~\cite{sverdlik2018phd}. At the beginning of a thesis, students often need help identifying suitable research topics~\cite{jilchaIdentifyingExistingResearch2025} and developing coherent proposals~\cite{fauzanUndergraduateStudentsDifficulties2022}, while supervisors must clarify expectations and provide feedback under time constraints. These challenges point to the need for preparatory processes that standardize early thesis supervision.

In response, the institute redesigned thesis preparation around an expos\'e-first workflow. The redesign built on workflow materials and review practices previously developed in a scientific-writing course to support expos\'e annotation, feedback exchange, and revision~\cite{zyskaPullRequestsClassroom2025}. However, adapting these materials and practices to thesis supervision for many parallel theses had not yet been described or examined, making this adaptation part of the present study's contribution. The workflow reconfigured roles: students became active authors of their research plans, supervisors shifted toward iterative feedback, feasibility checking, and preliminary quality assurance, administrators coordinated deadlines, submissions, communication, and platform transitions, and the professor provided feedback on topics before students began writing and on expos\'e drafts before final approval. This transformation made expectations and responsibilities more explicit, while increasing the importance of administrative coordination work that often remained in the background. Accordingly, we conceptualize the workflow not only as a pedagogical intervention, but as a sociotechnical transformation of interactions in academic supervision.

While prior work has mainly examined thesis supervision as a pedagogical relationship~\cite{grohnert2024effective, karunaratne2018blended}, this study focuses on the coordination work required to sustain thesis preparation across roles and the two digital platforms used in the workflow, Moodle and CARE\@. \textbf{RQ1} asks how students, supervisors, and administrators experience the expos\'e-first workflow as a structured process for early thesis preparation, and \textbf{RQ2} examines how the workflow redistributes responsibility, supervision, and administrative coordination work across roles and platforms.

To address these questions, we combine quantitative survey measures, including NASA-TLX workload ratings~\cite{HART1988139}, with qualitative interview data collected across two workflow iterations. We structure our analysis using the four reflective lenses proposed by Frauenberger et al.~\cite{frauenberger2015pursuit}: epistemology, values, stakeholders, and outcomes. These lenses allow us to examine how the workflow changes what counts as evidence of student understanding, which values shape the process, how responsibilities move across roles, and which outcomes depend on administrative coordination work.

This paper makes three contributions. First, we make explicit and analyze an expos\'e-first thesis workflow for many parallel thesis projects at a large research institute. Second, we show how the workflow redistributes responsibility across students, supervisors, administrators, and the professor. Third, we identify administrative coordination work as a central but underrecognized part of multi-platform thesis supervision.

\section{Related Work}

Research on thesis supervision has largely framed supervision as a pedagogical relationship between students and supervisors. Grohnert et al.~\cite{grohnert2024effective} synthesize prior work into an input--process--outcome framework focused on student and supervisor characteristics, relationships, actions, and outcomes. Technology-supported studies examine how supervision can be supported through blended supervision, ICT-enabled guidance, and peer interaction~\cite{karunaratne2018blended, aghaee2016ict}. Lagstedt et al.~\cite{lagstedt2020expertoriented} broaden this view by studying the digitalization of a university thesis process involving expert, coordination, and administrative roles. Together, this work shows the pedagogical and technical support needs of thesis supervision, but leaves room for analyzing how supervision workflows redistribute coordination work across stakeholders and institutional systems.

To analyze this coordination layer, we draw on CSCW work on articulation and invisible labor. Articulation work describes the effort required to support cooperative work arrangements in practice~\cite{schmidt1992taking}, while invisible labor highlights how work becomes recognized or overlooked depending on the indicators through which it is made visible~\cite{star1999layers}. Empirical studies further show how such work can appear in mediation between users and technologies~\cite{pallesenArticulationWorkMiddle2018} and in the hidden work required to make academic writing practices accessible to participants with different needs~\cite{wangInvisibleLaborAccess2022}. We use this lens to understand administrative coordination as work that compensates for infrastructural seams between platforms, where systems do not automatically exchange submissions, feedback, reminders, or approval status.

Finally, we treat the workflow as evolving infrastructure. Infrastructure is relational and often becomes visible in breakdown~\cite{star1996steps}, while infrastructuring research emphasizes the ongoing work of connecting technologies, practices, organizations, and stakeholders~\cite{karastiInfrastructuringParticipatoryDesign2014, bodker2017tyingknots, bilstrupMlmachineorgInfrastructuringResearch2024}. This motivates our use of Frauenberger et al.'s reflective lenses~\cite{frauenberger2015pursuit} to examine how the workflow redistributes responsibility, visibility, and coordination across stakeholder groups.

\section{Case Context and Workflow Design}
\label{sec:case-context}

Our study examines a thesis workflow redesign at a large European computer science research institute with more than 40 active researchers and approximately 20 theses per semester. Before the redesign, thesis preparation was handled largely on a case-by-case basis: students contacted the institute individually, supervisors proposed or shaped topics asynchronously, and administrative staff coordinated communication, documents, and approval steps through the central contact email. This reactive process made supervisory demand difficult to anticipate and produced uneven expectations across student--supervisor pairs.

The redesigned process introduced an \emph{expos\'e-first workflow}. The \emph{expos\'e} is a 3-page research proposal that students prepare before thesis registration. It describes the topic, motivation, research question, related work, planned method, and preliminary timeline. Previously, supervisors often wrote this document; now, students become its primary authors, while supervisors provide feedback and assess scientific feasibility. The thesis workflow adapts the draft--feedback--revision process used in the scientific-writing course, in which students submitted draft expos\'es, received structured feedback from peers and instructors, and revised their drafts into final research proposals~\cite{zyskaPullRequestsClassroom2025}. In the thesis workflow, the expos\'e functions both as a learning artifact and as a coordination artifact. It helps students develop an initial research plan while giving supervisors, administrators, and the professor a shared object for feedback, feasibility checking, and approval.

Figure~\ref{fig:workflow-overview} summarizes the redesigned workflow. It begins with \emph{topic announcement}, where supervisors prepare thesis topics and the professor reviews and approves them. Students may then attend an optional \emph{thesis fair} to learn about available topics before submitting an \emph{application}. After the application deadline, administrators review applications and coordinate \emph{topic matching}, which assigns students to topics and supervisors based on student preferences, topic fit, and available supervisory capacity. Matched students then enter the central \emph{expos\'e-writing} phase. During this phase, students draft and revise their expos\'e, while supervisors review drafts and provide feedback. The workflow ends with \emph{evaluation and approval}, where the professor provides feedback on expos\'e drafts and formally approves the final expos\'e before students enter the main thesis-writing phase.

The workflow distinguishes four stakeholder roles. \emph{Students} apply, develop the expos\'e, respond to feedback, and prepare the thesis project before registration. \emph{Supervisors} propose topics, participate in topic matching, and guide students through iterative feedback, feasibility checking, and preliminary quality assurance rather than writing the expos\'e themselves. \emph{Administrators} coordinate the process across deadlines, applications, communication channels, and digital platforms. The \emph{professor} reviews thesis topics, provides feedback on expos\'e drafts, and retains final approval and formal quality assurance authority for thesis topics and final expos\'es. Thus, the redesign does not remove work; it redistributes it across roles.

The workflow is supported by two disconnected digital platforms: Moodle, used for announcements and organizational communication, and CARE (Collaborative AI-Assisted Research Environment)~\cite{zyskaCARECollaborativeAIAssisted2023}, an established reviewing platform used for structured review and feedback. Because these systems are not integrated, administrators manually connect submissions, feedback, reminders, and approval steps across platforms. We refer to this as \emph{administrative coordination}: the background work required to keep the workflow moving across stakeholders and systems. This coordination role is central to our later analysis, because the workflow appears more structured at the participant-facing level while administrators handle the manual transfers, reminders, and status checks that make this structure possible.

\section{Methodology}
\label{sec:methodology}

We conducted an exploratory mixed-methods case study of the expos\'e-first thesis workflow in two summer semesters, 2025 (SS25) and 2026 (SS26). Data collection combined post-workflow surveys with semi-structured interviews. The surveys yielded 22 responses in total: 15 in SS25 (9 students, 6 supervisors) and 7 in SS26 (5 students, 2 supervisors). Overall, this corresponds to 14 student responses and 8 supervisor responses. The surveys captured perceived workflow clarity, feedback quality, role expectations, workload, and platform usability. Workload was measured using NASA-TLX, while platform usability was captured using UMUX~\cite{Kraig2010}.

To interpret the survey results, we conducted 9 interviews with students (STU), supervisors (SUP), and administrators (ADM). In SS25, we interviewed one student, three supervisors, and one administrator. In SS26, we interviewed one student, two supervisors, and one administrator who also held a supervisory role. Administrator perspectives are therefore based on interview data rather than survey responses. Each interview lasted approximately one hour. Recruitment was conducted via Moodle announcements and direct invitations, and all participation was voluntary. The interviews focused on participants' experiences with expos\'e writing and review, role responsibilities, feedback practices, timeline constraints, platform use, and coordination work.

\begin{figure*}[t]
  \centering

  \begin{subfigure}[t]{0.55\linewidth}
    \centering
    \includegraphics[width=\linewidth]{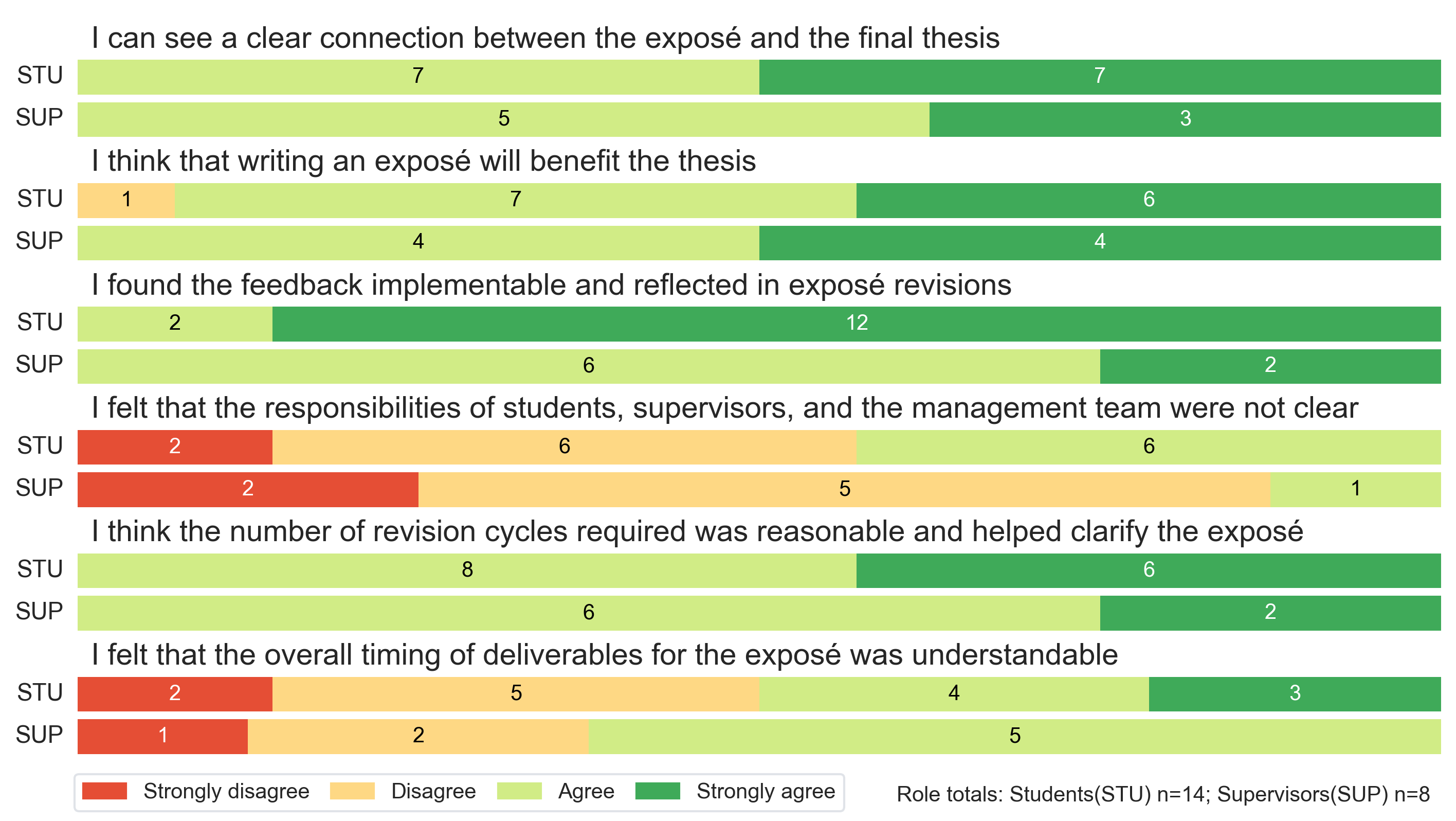}
    \caption{Workflow survey statements}
    \label{fig:workflow-stacked-bar}
  \end{subfigure}
  \hfill
  \begin{subfigure}[t]{0.44\linewidth}
    \centering
    \includegraphics[width=\linewidth]{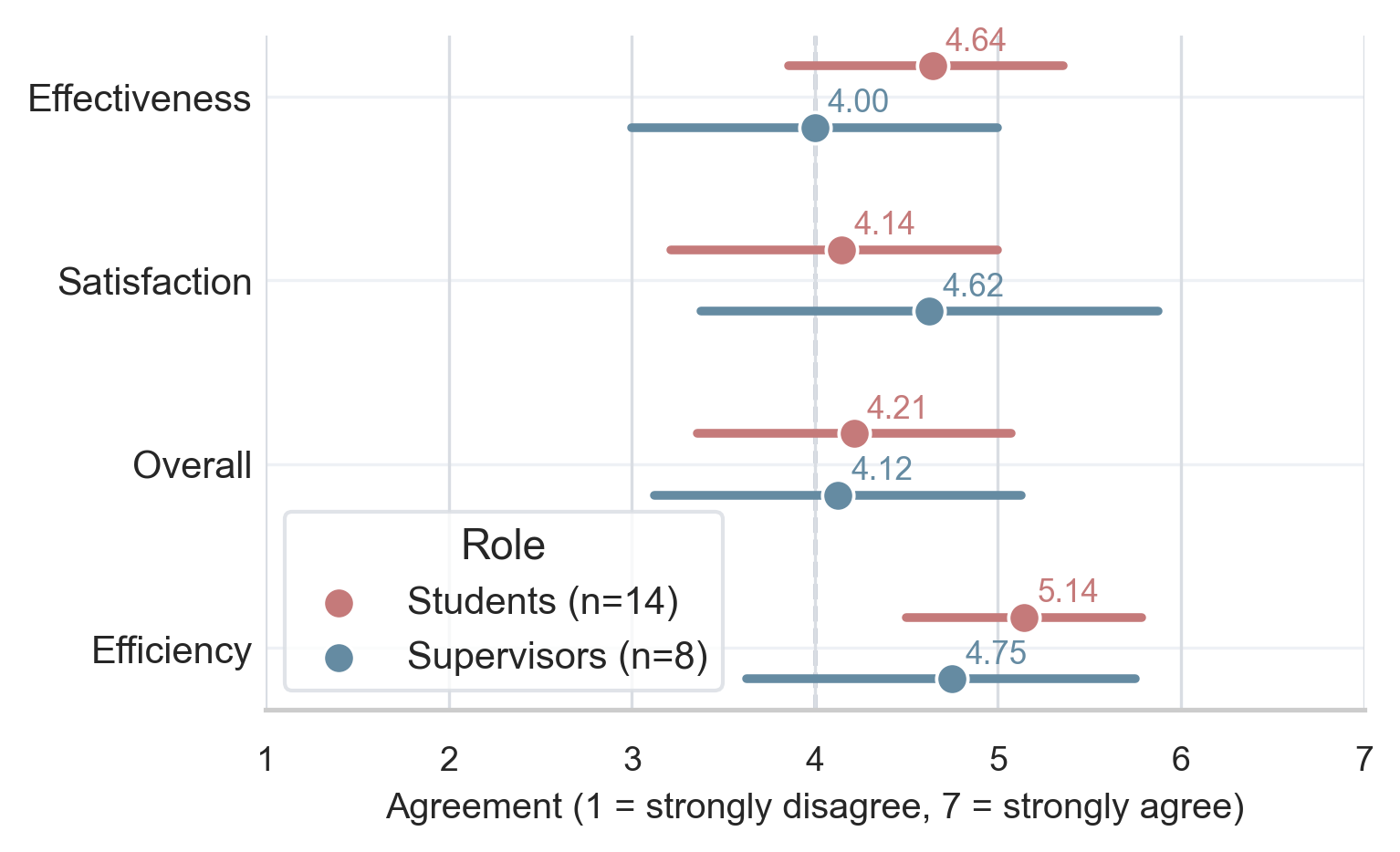}
    \caption{UMUX scores}
    \label{fig:umux-graph}
  \end{subfigure}

  \caption{Survey results on the exposé workflow and perceived usability. Panel~(a) shows students' and supervisors' agreement with statements about exposé-thesis connection, feedback implementation, revision cycles, responsibilities, and timing. Panel~(b) reports UMUX scores for effectiveness, satisfaction, overall usability, and efficiency, comparing students and supervisors.}
  \Description{A two-panel figure. The left panel compares student and supervisor responses to workflow statements; the right panel compares UMUX usability scores across effectiveness, satisfaction, overall usability, and efficiency.}
  \label{fig:umux-row}
\end{figure*}

Survey data were analyzed descriptively to identify patterns across stakeholder roles and workflow iterations. Interview transcripts were analyzed thematically. After identifiers were removed, one doctoral researcher and two student researchers independently coded the transcripts. We used Frauenberger et al.'s four lenses as a deductive analytic framework, and developed themes inductively within each lens. The three coders then discussed coding differences and resolved them through consensus, forming a cross-transcript thematic grouping. We integrated the survey and interview findings by using the quantitative results to identify patterns and the interview data to explain how these patterns emerged in practice.

\section{Findings and Discussion}
\label{sec:findings-discussion}

The four subsections follow Frauenberger et al.'s reflective lenses: epistemology concerns knowledge production, values concerns design priorities and tensions, stakeholders concerns whose roles and perspectives are foregrounded, and outcomes concerns intended and unintended consequences in practice.

\paragraph*{\textbf{Epistemology: Expos\'e authorship made learning visible, but not fully measurable.}}

The workflow rests on the assumption that students learn by authoring an expos\'e themselves. One supervisor explained that ``if you just write it for them, then they will not learn about all this background work and the project'' (SUP2). Students similarly described the expos\'e as a way of understanding the overall thesis trajectory: ``because I did the expos\'e where you have to think about all of the thesis, then I know what's coming now'' (STU1). Survey results show a similar pattern (Figure~\ref{fig:umux-row}\subref{fig:workflow-stacked-bar}). All students reported a clear connection between the exposé and the final thesis (14/14; M = 3.50, 4-point scale). Supervisors showed a similar pattern, with all respondents agreeing with the statement from their perspective (8/8; M = 3.38, 4-point scale). Yet, the written expos\'e captured only part of the learning process, which also unfolds through oral discussion, clarification, and revision. The artifact is therefore educationally meaningful, but not a complete record of learning.

Because the workflow treats the written expos\'e as evidence of early research planning and understanding, reported AI use in SS26 became relevant to how this artifact could be interpreted. All five student respondents reported using AI for paraphrasing or rewriting, while supervisors used it primarily to assess writing quality or potential AI use. One supervisor noted that he ``had the feeling that the student is responsible in using AI,'' while also acknowledging that ``from just the expos\'e, I can't really tell how much of that is just the AI and how much is actually the student, but the result in quality was very high'' (SUP1). A student also described AI as a revision aid: she first drafted sentences herself, then asked AI to ``make it a bit better,'' but revised outputs that were ``not what I want'' (STU1). The expos\'e thus remained a valuable learning artifact, but its written quality became less reliable as evidence of individual understanding.

\paragraph*{\textbf{Values: Structure improved preparation while creating new value tensions.}}
The workflow improved structure and expectation alignment, although these effects were not uniform across all items. As shown in Figure~\ref{fig:umux-row}\subref{fig:workflow-stacked-bar}, all student respondents reported that supervisor feedback was implementable in their revisions (14/14; $M = 3.86$, 4-point scale). Most respondents also agreed that writing an expos\'e would benefit the thesis, with high agreement among both students ($M = 3.36$, 4-point scale) and supervisors ($M = 3.50$, 4-point scale). Workload ratings suggest that students experienced the workflow as more demanding than supervisors. This difference was visible in ratings of mental demand (65.00 vs.\ 58.75), effort (75.36 vs.\ 65.00), and frustration (54.64 vs.\ 47.50), with values reported as students versus supervisors throughout. Interview data particularly points to time pressure as part of the burden. One student stated that ``I would probably start looking for it after my exams. But at this time, I didn't even finish my last exam and the topic selection was during exams'' (STU1).

The workflow also introduced value tensions. Student authorship supported learning and preparation, but it also shifted work away from supervisors and onto students. The fixed timeline created predictability, but it also constrained the time available for proposal development. Finally, formal access through topic announcement and topic matching did not fully remove inequalities caused by prior knowledge, timing, or informal connections. As one student noted about the short submission deadline, without prior involvement in the lab, she ``would have missed it probably too'' (STU1).

\paragraph*{\textbf{Stakeholders: Responsibility shifted visibly to students and invisibly to administrators.}}
The most visible role shift was from supervisors to students. Students became responsible for drafting the expos\'e, while supervisors moved from proposal authorship to iterative feedback, feasibility checking, and preliminary quality assurance. This did not simply reduce supervisory work. Some supervisors reported that repeated feedback rounds required as much effort as writing the expos\'e themselves; one stated that the ``time of efforts was doubled compared to the old workflow where I wrote the expos\'e myself'' (SUP3). The redesign therefore changed the kind of supervisory work rather than eliminating it.

The less visible shift concerned administrative labor. Administrators coordinated applications, matching, deadlines, reminders, document transfer, and communication across platforms. One administrator described this role as being ``basically the conductor of the whole process'' (ADM1). Standardization for students and supervisors was partly achieved through intensified background coordination by administrators. Across the two semester-long workflow iterations, SS25 and SS26, administrators coordinated 100 applications, 39 thesis students, 31 supervisors, and 36 thesis topics. This coordination also became more concentrated across iterations, as the administrative team decreased from three people in SS25 to one in SS26. The workflow thus created articulation work that remained visible mainly when it failed~\cite{star1999layers}.

\paragraph*{\textbf{Outcomes: Standardization worked, but the infrastructure remained fragile.}}

Overall, the workflow standardized thesis preparation. Students gained better orientation, clearer expectations, and a stronger basis for thesis writing. The exposé became a shared artifact that students returned to. CARE supported this process when its feedback features matched supervisory practice; one student valued the color-coded annotations because they distinguished wording issues from global feedback (STU1).

Yet the workflow remained infrastructurally fragile.
As shown in Figure~\ref{fig:umux-row}\subref{fig:umux-graph}, standard UMUX composite scores, transformed to a 0–100 scale, indicated persistent usability friction with CARE. Students reported higher perceived alignment with their requirements ($M = 58.93$, $SD = 22.22$, $N = 14$) than supervisors ($M = 56.25$, $SD = 23.88$, $N = 8$).
This interpretation is consistent with qualitative feedback: students also pointed to the burden of using multiple systems, with one summarizing the issue as: ``I would like only one platform'' (STU1). Because the learning-management system and CARE were not integrated, administrators manually connected submissions, feedback, reminders, and approval steps. The workflow therefore redistributed responsibility faster than it built the infrastructure needed to support that redistribution.

\section{Conclusion}
\label{sec:conclusion}
This study shows that the expos\'e-first thesis workflow was perceived by students and supervisors as making thesis preparation more structured, transparent, and pedagogically useful. However, the redesign did not reduce work overall; it redistributed it. Students assumed authorship earlier, supervisors shifted toward iterative feedback, feasibility checking, and preliminary quality assurance, and administrators carried much of the coordination required to bridge disconnected platforms, deadlines, submissions, and feedback processes.

The study is limited by its small scale, single-institution setting, and its focus on perceived workflow experiences rather than long-term thesis outcomes. Because the workflow was designed and studied within the same institutional context, participants may have been reluctant to criticize a process associated with their own institute; positive evaluations should therefore be interpreted in light of possible social desirability effects. Even so, the case shows that standardizing thesis preparation can improve orientation and expectation alignment while shifting coordination work to administrators. Future research should examine how such workflows transfer across institutions and how they affect thesis quality, student learning, and sustainable supervisory capacity.

\section{Ethical Considerations}
\label{sec:ethics}

This study was approved by the University's Institutional Review Board under IRB number EK 56/2025. All participants provided informed consent for the collection, analysis, and reporting of aggregated findings and anonymized quotations. Participation was voluntary, and transcripts were anonymized before analysis. Generative AI tools were used for language editing and manuscript refinement; all analysis and interpretations were developed, reviewed, and approved by the authors.

\begin{acks}
This work has received funding from the European Union (ERC, InterText, 101054961). Views and opinions expressed are however those of the author(s) only and do not necessarily reflect those of the European Union or the European Research Council. Neither the European Union nor the granting authority can be held responsible for them.
\end{acks}

\bibliographystyle{ACM-Reference-Format}
\bibliography{references}


\begin{thebibliography}{21}


\ifx \showCODEN    \undefined \def \showCODEN     #1{\unskip}     \fi
\ifx \showDOI      \undefined \def \showDOI       #1{#1}\fi
\ifx \showISBNx    \undefined \def \showISBNx     #1{\unskip}     \fi
\ifx \showISBNxiii \undefined \def \showISBNxiii  #1{\unskip}     \fi
\ifx \showISSN     \undefined \def \showISSN      #1{\unskip}     \fi
\ifx \showLCCN     \undefined \def \showLCCN      #1{\unskip}     \fi
\ifx \shownote     \undefined \def \shownote      #1{#1}          \fi
\ifx \showarticletitle \undefined \def \showarticletitle #1{#1}   \fi
\ifx \showURL      \undefined \def \showURL       {\relax}        \fi
\providecommand\bibfield[2]{#2}
\providecommand\bibinfo[2]{#2}
\providecommand\natexlab[1]{#1}
\providecommand\showeprint[2][]{arXiv:#2}

\bibitem[Aghaee and Keller(2016)]%
        {aghaee2016ict}
\bibfield{author}{\bibinfo{person}{Naghmeh Aghaee} {and}
  \bibinfo{person}{Christina Keller}.} \bibinfo{year}{2016}\natexlab{}.
\newblock \showarticletitle{{ICT}-Supported Peer Interaction among Learners in
  {Bachelor's} and {Master's} Thesis Courses}.
\newblock \bibinfo{journal}{\emph{Computers \& Education}}
  \bibinfo{volume}{94} (\bibinfo{year}{2016}), \bibinfo{pages}{276--297}.
\newblock
\urldef\tempurl%
\url{https://doi.org/10.1016/j.compedu.2015.11.006}
\showDOI{\tempurl}


\bibitem[Bilstrup et~al\mbox{.}(2024)]%
        {bilstrupMlmachineorgInfrastructuringResearch2024}
\bibfield{author}{\bibinfo{person}{Karl-Emil~Kj{\ae}r Bilstrup},
  \bibinfo{person}{Magnus~H{\o}holt Kaspersen}, \bibinfo{person}{Niels~Olof
  Bouvin}, {and} \bibinfo{person}{Marianne~Graves Petersen}.}
  \bibinfo{year}{2024}\natexlab{}.
\newblock \showarticletitle{Ml-Machine.Org: {{Infrastructuring}} a {{Research
  Product}} to {{Disseminate AI Literacy}} in {{Education}}}. In
  \bibinfo{booktitle}{\emph{Proceedings of the {{CHI Conference}} on {{Human
  Factors}} in {{Computing Systems}}}}. \bibinfo{publisher}{ACM},
  \bibinfo{address}{Honolulu HI USA}, \bibinfo{pages}{1--16}.
\newblock
\showISBNx{979-8-4007-0330-0}
\urldef\tempurl%
\url{https://doi.org/10.1145/3613904.3642539}
\showDOI{\tempurl}


\bibitem[B{\o}dker et~al\mbox{.}(2017)]%
        {bodker2017tyingknots}
\bibfield{author}{\bibinfo{person}{Susanne B{\o}dker},
  \bibinfo{person}{Christian Dindler}, {and} \bibinfo{person}{Ole~Sejer
  Iversen}.} \bibinfo{year}{2017}\natexlab{}.
\newblock \showarticletitle{Tying Knots: Participatory Infrastructuring at
  Work}.
\newblock \bibinfo{journal}{\emph{Computer Supported Cooperative Work}}
  \bibinfo{volume}{26}, \bibinfo{number}{1-2} (\bibinfo{year}{2017}),
  \bibinfo{pages}{245--273}.
\newblock
\urldef\tempurl%
\url{https://doi.org/10.1007/s10606-017-9268-y}
\showDOI{\tempurl}


\bibitem[Fauzan et~al\mbox{.}(2022)]%
        {fauzanUndergraduateStudentsDifficulties2022}
\bibfield{author}{\bibinfo{person}{Umar Fauzan}, \bibinfo{person}{Nur Hasanah},
  {and} \bibinfo{person}{Siti Hadijah}.} \bibinfo{year}{2022}\natexlab{}.
\newblock \showarticletitle{The {{Undergraduate Students}}' {{Difficulties}} in
  {{Writing Thesis Proposal}}}.
\newblock \bibinfo{journal}{\emph{Indonesian Journal of EFL and Linguistics}}
  \bibinfo{volume}{7}, \bibinfo{number}{1} (\bibinfo{date}{May}
  \bibinfo{year}{2022}), \bibinfo{pages}{175}.
\newblock
\showISSN{2503-4197, 2527-5070}
\urldef\tempurl%
\url{https://doi.org/10.21462/ijefl.v7i1.515}
\showDOI{\tempurl}


\bibitem[Finstad(2010)]%
        {Kraig2010}
\bibfield{author}{\bibinfo{person}{Kraig Finstad}.}
  \bibinfo{year}{2010}\natexlab{}.
\newblock \showarticletitle{The Usability Metric for User Experience}.
\newblock \bibinfo{journal}{\emph{Interacting with Computers}}
  \bibinfo{volume}{22}, \bibinfo{number}{5} (\bibinfo{date}{09}
  \bibinfo{year}{2010}), \bibinfo{pages}{323--327}.
\newblock
\showISSN{0953-5438}
\urldef\tempurl%
\url{https://doi.org/10.1016/j.intcom.2010.04.004}
\showDOI{\tempurl}
\showeprint{https://academic.oup.com/iwc/article-pdf/22/5/323/1992916/iwc22-0323.pdf}


\bibitem[Frauenberger et~al\mbox{.}(2015)]%
        {frauenberger2015pursuit}
\bibfield{author}{\bibinfo{person}{Christopher Frauenberger},
  \bibinfo{person}{Judith Good}, \bibinfo{person}{Geraldine Fitzpatrick}, {and}
  \bibinfo{person}{Ole~Sejer Iversen}.} \bibinfo{year}{2015}\natexlab{}.
\newblock \showarticletitle{In pursuit of rigour and accountability in
  participatory design}.
\newblock \bibinfo{journal}{\emph{International Journal of Human-Computer
  Studies}}  \bibinfo{volume}{74} (\bibinfo{year}{2015}),
  \bibinfo{pages}{93--106}.
\newblock
\urldef\tempurl%
\url{https://doi.org/10.1016/j.ijhcs.2014.09.004}
\showDOI{\tempurl}


\bibitem[Grohnert et~al\mbox{.}(2024)]%
        {grohnert2024effective}
\bibfield{author}{\bibinfo{person}{Therese Grohnert}, \bibinfo{person}{Lena
  Gromotka}, \bibinfo{person}{Inken Gast}, \bibinfo{person}{Laurie Delnoij},
  {and} \bibinfo{person}{Simon Beausaert}.} \bibinfo{year}{2024}\natexlab{}.
\newblock \showarticletitle{Effective master's thesis supervision -- A
  summative framework for research and practice}.
\newblock \bibinfo{journal}{\emph{Educational Research Review}}
  \bibinfo{volume}{42} (\bibinfo{year}{2024}), \bibinfo{pages}{100589}.
\newblock
\urldef\tempurl%
\url{https://doi.org/10.1016/j.edurev.2023.100589}
\showDOI{\tempurl}


\bibitem[Hart and Staveland(1988)]%
        {HART1988139}
\bibfield{author}{\bibinfo{person}{Sandra~G. Hart} {and}
  \bibinfo{person}{Lowell~E. Staveland}.} \bibinfo{year}{1988}\natexlab{}.
\newblock \showarticletitle{Development of NASA-TLX (Task Load Index): Results
  of Empirical and Theoretical Research}.
\newblock In \bibinfo{booktitle}{\emph{Human Mental Workload}},
  \bibfield{editor}{\bibinfo{person}{Peter~A. Hancock} {and}
  \bibinfo{person}{Najmedin Meshkati}} (Eds.). \bibinfo{series}{Advances in
  Psychology}, Vol.~\bibinfo{volume}{52}. \bibinfo{publisher}{North-Holland},
  \bibinfo{address}{Amsterdam}, \bibinfo{pages}{139--183}.
\newblock
\showISSN{0166-4115}
\urldef\tempurl%
\url{https://doi.org/10.1016/S0166-4115(08)62386-9}
\showDOI{\tempurl}


\bibitem[Jilcha(2025)]%
        {jilchaIdentifyingExistingResearch2025}
\bibfield{author}{\bibinfo{person}{Kassu Jilcha}.}
  \bibinfo{year}{2025}\natexlab{}.
\newblock \showarticletitle{Identifying Existing Research Challenges and
  Enhancing Outcomes through the Development of Standardized Methodologies}.
\newblock \bibinfo{journal}{\emph{Humanities and Social Sciences
  Communications}} \bibinfo{volume}{12}, \bibinfo{number}{1}
  (\bibinfo{date}{Feb.} \bibinfo{year}{2025}), \bibinfo{pages}{187}.
\newblock
\showISSN{2662-9992}
\urldef\tempurl%
\url{https://doi.org/10.1057/s41599-024-04269-7}
\showDOI{\tempurl}


\bibitem[Karasti(2014)]%
        {karastiInfrastructuringParticipatoryDesign2014}
\bibfield{author}{\bibinfo{person}{Helena Karasti}.}
  \bibinfo{year}{2014}\natexlab{}.
\newblock \showarticletitle{Infrastructuring in Participatory Design}. In
  \bibinfo{booktitle}{\emph{Proceedings of the 13th {{Participatory Design
  Conference}} on {{Research Papers}} - {{PDC}} '14}}. \bibinfo{publisher}{ACM
  Press}, \bibinfo{address}{Windhoek, Namibia}, \bibinfo{pages}{141--150}.
\newblock
\showISBNx{978-1-4503-2256-0}
\urldef\tempurl%
\url{https://doi.org/10.1145/2661435.2661450}
\showDOI{\tempurl}


\bibitem[Karunaratne(2018)]%
        {karunaratne2018blended}
\bibfield{author}{\bibinfo{person}{Thashmee Karunaratne}.}
  \bibinfo{year}{2018}\natexlab{}.
\newblock \showarticletitle{Blended Supervision for Thesis Projects in Higher
  Education: A Case Study}.
\newblock \bibinfo{journal}{\emph{Electronic Journal of e-Learning}}
  \bibinfo{volume}{16}, \bibinfo{number}{2} (\bibinfo{year}{2018}),
  \bibinfo{pages}{79--90}.
\newblock


\bibitem[Lagstedt et~al\mbox{.}(2020)]%
        {lagstedt2020expertoriented}
\bibfield{author}{\bibinfo{person}{Altti Lagstedt}, \bibinfo{person}{Juha~P.
  Lindstedt}, {and} \bibinfo{person}{Raine Kauppinen}.}
  \bibinfo{year}{2020}\natexlab{}.
\newblock \showarticletitle{An Outcome of Expert-Oriented Digitalization of
  University Processes}.
\newblock \bibinfo{journal}{\emph{Education and Information Technologies}}
  \bibinfo{volume}{25} (\bibinfo{year}{2020}), \bibinfo{pages}{5853--5871}.
\newblock
\urldef\tempurl%
\url{https://doi.org/10.1007/s10639-020-10252-x}
\showDOI{\tempurl}


\bibitem[Pallesen and Jacobsen(2018)]%
        {pallesenArticulationWorkMiddle2018}
\bibfield{author}{\bibinfo{person}{Trine Pallesen} {and}
  \bibinfo{person}{Peter~H. Jacobsen}.} \bibinfo{year}{2018}\natexlab{}.
\newblock \showarticletitle{Articulation Work from the Middle---a Study of How
  Technicians Mediate Users and Technology}.
\newblock \bibinfo{journal}{\emph{New Technology, Work and Employment}}
  \bibinfo{volume}{33}, \bibinfo{number}{2} (\bibinfo{year}{2018}),
  \bibinfo{pages}{171--186}.
\newblock
\urldef\tempurl%
\url{https://doi.org/10.1111/ntwe.12113}
\showDOI{\tempurl}


\bibitem[Schmidt and Bannon(1992)]%
        {schmidt1992taking}
\bibfield{author}{\bibinfo{person}{Kjeld Schmidt} {and} \bibinfo{person}{Liam
  Bannon}.} \bibinfo{year}{1992}\natexlab{}.
\newblock \showarticletitle{Taking {CSCW} Seriously: Supporting Articulation
  Work}.
\newblock \bibinfo{journal}{\emph{Computer Supported Cooperative Work (CSCW)}}
  \bibinfo{volume}{1}, \bibinfo{number}{1--2} (\bibinfo{year}{1992}),
  \bibinfo{pages}{7--40}.
\newblock
\urldef\tempurl%
\url{https://doi.org/10.1007/BF00752449}
\showDOI{\tempurl}


\bibitem[Star and Ruhleder(1996)]%
        {star1996steps}
\bibfield{author}{\bibinfo{person}{Susan~Leigh Star} {and}
  \bibinfo{person}{Karen Ruhleder}.} \bibinfo{year}{1996}\natexlab{}.
\newblock \showarticletitle{Steps Toward an Ecology of Infrastructure: Design
  and Access for Large Information Spaces}.
\newblock \bibinfo{journal}{\emph{Information Systems Research}}
  \bibinfo{volume}{7}, \bibinfo{number}{1} (\bibinfo{year}{1996}),
  \bibinfo{pages}{111--134}.
\newblock
\urldef\tempurl%
\url{https://doi.org/10.1287/isre.7.1.111}
\showDOI{\tempurl}


\bibitem[Star and Strauss(1999)]%
        {star1999layers}
\bibfield{author}{\bibinfo{person}{Susan~Leigh Star} {and}
  \bibinfo{person}{Anselm Strauss}.} \bibinfo{year}{1999}\natexlab{}.
\newblock \showarticletitle{Layers of Silence, Arenas of Voice: The Ecology of
  Visible and Invisible Work}.
\newblock \bibinfo{journal}{\emph{Computer Supported Cooperative Work (CSCW)}}
  \bibinfo{volume}{8}, \bibinfo{number}{1--2} (\bibinfo{year}{1999}),
  \bibinfo{pages}{9--30}.
\newblock
\urldef\tempurl%
\url{https://doi.org/10.1023/A:1008651105359}
\showDOI{\tempurl}


\bibitem[Sverdlik et~al\mbox{.}(2018)]%
        {sverdlik2018phd}
\bibfield{author}{\bibinfo{person}{Anna Sverdlik}, \bibinfo{person}{Nathan~C.
  Hall}, \bibinfo{person}{Lynn McAlpine}, {and} \bibinfo{person}{Kent
  Hubbard}.} \bibinfo{year}{2018}\natexlab{}.
\newblock \showarticletitle{The {PhD} Experience: A Review of the Factors
  Influencing Doctoral Students' Completion, Achievement, and Well-Being}.
\newblock \bibinfo{journal}{\emph{International Journal of Doctoral Studies}}
  \bibinfo{volume}{13} (\bibinfo{year}{2018}), \bibinfo{pages}{361--388}.
\newblock
\urldef\tempurl%
\url{https://doi.org/10.28945/4113}
\showDOI{\tempurl}


\bibitem[{UNESCO}(2025)]%
        {unesco2025higher}
\bibfield{author}{\bibinfo{person}{{UNESCO}}.} \bibinfo{year}{2025}\natexlab{}.
\newblock \bibinfo{booktitle}{\emph{Higher Education: Figures at a Glance}}.
\newblock \bibinfo{type}{{T}echnical {R}eport} ED/PLS/HED/2025/02.
  \bibinfo{institution}{UNESCO}.
\newblock


\bibitem[Wang and Piper(2022)]%
        {wangInvisibleLaborAccess2022}
\bibfield{author}{\bibinfo{person}{Emily~Q. Wang} {and}
  \bibinfo{person}{Anne~Marie Piper}.} \bibinfo{year}{2022}\natexlab{}.
\newblock \showarticletitle{The {{Invisible Labor}} of {{Access}} in {{Academic
  Writing Practices}}: {{A Case Analysis}} with {{Dyslexic Adults}}}.
\newblock \bibinfo{journal}{\emph{Proceedings of the ACM on Human-Computer
  Interaction}} \bibinfo{volume}{6}, \bibinfo{number}{CSCW1}
  (\bibinfo{date}{March} \bibinfo{year}{2022}), \bibinfo{pages}{1--25}.
\newblock
\showISSN{2573-0142}
\urldef\tempurl%
\url{https://doi.org/10.1145/3512967}
\showDOI{\tempurl}


\bibitem[Zyska et~al\mbox{.}(2023)]%
        {zyskaCARECollaborativeAIAssisted2023}
\bibfield{author}{\bibinfo{person}{Dennis Zyska}, \bibinfo{person}{Nils Dycke},
  \bibinfo{person}{Jan Buchmann}, \bibinfo{person}{Ilia Kuznetsov}, {and}
  \bibinfo{person}{Iryna Gurevych}.} \bibinfo{year}{2023}\natexlab{}.
\newblock \showarticletitle{{{CARE}}: {{Collaborative AI-Assisted Reading
  Environment}}}. In \bibinfo{booktitle}{\emph{Proceedings of the 61st {{Annual
  Meeting}} of the {{Association}} for {{Computational Linguistics}}
  ({{Volume}} 3: {{System Demonstrations}})}},
  \bibfield{editor}{\bibinfo{person}{Danushka Bollegala},
  \bibinfo{person}{Ruihong Huang}, {and} \bibinfo{person}{Alan Ritter}} (Eds.).
  \bibinfo{publisher}{Association for Computational Linguistics},
  \bibinfo{address}{Toronto, Canada}, \bibinfo{pages}{291--303}.
\newblock
\urldef\tempurl%
\url{https://doi.org/10.18653/v1/2023.acl-demo.28}
\showDOI{\tempurl}


\bibitem[Zyska et~al\mbox{.}(2025)]%
        {zyskaPullRequestsClassroom2025}
\bibfield{author}{\bibinfo{person}{Dennis Zyska}, \bibinfo{person}{Ilia
  Kuznetsov}, \bibinfo{person}{Florian M{\"u}ller}, {and}
  \bibinfo{person}{Iryna Gurevych}.} \bibinfo{year}{2025}\natexlab{}.
\newblock \showarticletitle{Pull {{Requests From The Classroom}}:
  {{Co-Developing Curriculum And Code}}}. In
  \bibinfo{booktitle}{\emph{Proceedings of the {{Mensch}} Und {{Computer}}
  2025}}. \bibinfo{publisher}{ACM}, \bibinfo{address}{Chemnitz Germany},
  \bibinfo{pages}{689--693}.
\newblock
\showISBNx{979-8-4007-1582-2}
\urldef\tempurl%
\url{https://doi.org/10.1145/3743049.3748581}
\showDOI{\tempurl}


\end{thebibliography}
\end{document}